\documentclass[a4paper,11pt,titlepage]{article}

\usepackage[latin1]{inputenc}
 \usepackage[T1]{fontenc}
 \usepackage[normalem]{ulem}
 \usepackage[english]{babel}
 \usepackage{verbatim}
 \usepackage{graphicx}
\usepackage{algorithmic} 
\usepackage{algorithm} 
\usepackage{amsthm, amsmath, amssymb, amsfonts}
\usepackage[modulo,mathlines]{lineno}
\usepackage{array} 
\usepackage{hyperref}

\usepackage{chngpage}
\usepackage[round]{natbib}
\usepackage{amssymb,amsmath,amsthm,amscd}
\usepackage{mathrsfs,IEEEtrantools}
\usepackage[left=1.5in, right=1.0in, top=2.0in, bottom=1.0in]{geometry}
\theoremstyle{plain} 
\newtheorem{theo}{Theorem}

\newtheorem{prop}[theo]{Proposition}

\usepackage{lmodern}
\rmfamily
\DeclareFontShape{T1}{lmr}{b}{sc}{<->ssub*cmr/bx/sc}{}
\DeclareFontShape{T1}{lmr}{bx}{sc}{<->ssub*cmr/bx/sc}{}

\begin{document}

\vspace{2 cm} 
\title{S|S|M:\\ Inference for time series analysis with State Space Models\\ \vspace{1 cm} {\large \texttt{ cat theta.json | ./ksimplex | ./kmcmc | ./pmcmc}}}

\author{Joseph Dureau, S\'ebastien Ballesteros, Tiffany Bogich}

\maketitle

\section{Introduction}

The main motivation behind the open source library \emph{SSM} is to reduce the technical friction that prevents modellers from sharing their work, quickly iterating in crisis situations, and making their work directly usable by public authorities to serve decision-making.
An illustration of this problem is the fact that the scientific production on the 2009 H1N1 epidemic has peaked three years after, in  2012, and that the vast majority of this production takes the form of static  \texttt{pdf} files. Even when the corresponding code is shared, it takes a substantial amount of time and effort for any other person than the author to at least reproduce the published results. 

This document gathers the methodological aspects on which the \emph{SSM} library is built on. Its first purpose is to provide full transparency on the general modeling framework that is proposed, and more precisely which are the available mathematical formulations for these models. It also describes the different inference algorithms implemented in the library, that are based on the state of the art of computational statistics.  All the source code is open and available on github (\href{https://github.com/standard-analytics/ssm}{https://github.com/standard-analytics/ssm}), and all contributions, comments and suggestions are more than welcome!

The modelling facet of the library stemmed from a focus on epidemiology and ecology, but it is more widely targeted to all systems that can be represented as c systems of ordinary or stochastic differential equations, compartmental models and combinations thereof.  While we refer the reader to the classic literature for ordinary and stochastic differential equations, the modelling framework for compartmental models is presented in Section \ref{sec:models} of this document, that proposes a grammar from which models can be  defined in a simple and non-ambiguous way, freeing their fundamental description from their mathematical formulation and technical implementation. Stochastic differential equations are simply described through their deterministic drift and dispersion matrix. This definition of compartmental models simply relies on the definition of transformations of the system, called reactions, that occur at given rates. We extend this classic perspective with the possibility of introducing environmental stochasticity, that captures potential mis-specifications of the model. The later is based on the propositions made in \cite{Breto2009}.

This mathematical formalisms available in this library to describe state space models are specially suited to particle-based methods as iterated filtering \citep{Ionides2011} and the particle MCMC \citep{Andrieu2010}. However, these methods remain computationally intensive which limits their applicability. Following the approach proposed in \cite{Dureau2013a}, the \emph{SSM} library also proposes algorithms based on the Extended Kalman Filter,  that allow drastic cost reductions in preliminary explorations of targeted densities and efficient initialisations of particle-based methods. These aspects will be described in Section \ref{sec:methods}.

At last, we provide in Section \ref{sec:applications}  three illustrations of how compartmental models and inference methods can be used to explore historic datasets, monitor current epidemics, and forecast their future evolution.

\section{\label{sec:models}Compartmental models}
\subsection{Definition}

Compartmental models are a general framework used to represent the state of a countable population (of humans, animals, molecules, etc) and its evolution. At a given time $t$, the population is described by the number of individuals in each of $c$ possible states: the ensemble of individuals in a same state defines what is termed as a compartment. Each individual belongs to one and only one compartment. Individuals within a same compartment are considered indistinguishable. We will consider in this document that $c$ is known, fixed, and finite. 

Each compartment can correspond to very diverse characterisations, depending on the context. They can be used to track the status of an individual with regards to a given disease in a human or animal population (susceptible or infected, for example), their age and/or their geographical location \citep{Anderson1992}. Additionnally, compartments can be used to track the number of specimens of different animal species in an ecosystem, as in the Lotka-Volterra predator-prey model \citep{Pulliam1988}. They can also be used in physics and chemistry to characterise molecule types, electronic charge or radioactive states \citep{Nagashima1968,Zanzonico2000}. Less classical illustrations of the use of compartmental models include tracking the spread of rumors among a population,  spread of obesity, or the propagation of economic difficulties among countries following a financial crisis \citep{Morris1993,Demiris2012}.

We note $z^{(i)}_t$ the size of compartment $i$ ($1\leq i \leq c$) at time $t$, and $z_t=[z_t^{(1)},..,z_t^{(c)}]$. A model is defined by a (finite) number $m$ of transformations of the system called reactions (the ensemble of all indexes is noted $\mathcal{R})$. These reactions correspond to one or several individuals passing from one compartment to another, or arriving or leaving the total population. In any case, each reaction $k$ is characterised by its effect on the structure of the population corresponding to a vector $l^{(k)}\in\mathbb{Z}^c$, and its intensity of occurrence. In the remainder of this document, we will make the classic assumption that the probability of occurrence of each reaction is proportional to the number of individuals in a given compartment (for individuals coming from outside the population of interest, a artificial source state can be introduced). This property can be specified through a mapping $\chi:\mathcal{R}\rightarrow [1;c]$ such that the transition rate of reaction $k$ can be written $r^{(k)}_t(z_t,\theta) z^{\chi(k)}_t$. This assumption implies the density-dependance of transition rates, i.e. the transition rates of the model where the state variable has been normalised ($\dot{z}_t=z_t/N$) can be simply written as $r^{(k)}_t(z_t,\theta) \dot{z}^{\chi(k)}_t$.

We allow for these rates to depend on time, in order to reflect  potential variations of external drivers of the system. They depend on a finite set of constant quantities gathered in a parameter vector $\theta$. In the remaining of this document, we will define compartmental models using the following formalism:

\begin{center}
\begin{tabular}{ccc}
\textbf{Reaction} & \textbf{Effect}   & \textbf{Rate}   \tabularnewline
\hline
reaction $1$ & $z_t\rightarrow z_t + l^{(1)}$ & $r^{(1)}(z_t,\theta)$  \tabularnewline
... 		& ... & ...  \tabularnewline
reaction $k$  & $z_t\rightarrow z_t + l^{(k)}$ & $r^{(k)}(z_t,\theta)$  \tabularnewline
... 		& ... & ...  \tabularnewline
reaction  $m$ & $z_t\rightarrow z_t + l^{(m)}$ & $r^{(m)}(z_t,\theta)$  \tabularnewline
\end{tabular}
\end{center}

Formally, this framework leads to the definition of a Markovian jump process, which dynamic can be expressed 
in the following way:

\begin{center}
\underline{Markovian jump process compartmental model}
\begin{IEEEeqnarray}{rCl}
\label{eq:ReferenceJump}
P(z_{t+dt}=z_t+l^{(k)}|z_t) &=&r_t^{(k)}(z_t,\theta) z^{\chi(k)}_t dt + o(dt) \;\;\;\;\;\ \text{for any } k\in \mathcal{R}\\
P(z_{t+dt}=z_t|z_t) &=& \big(1-\sum_{ k\in \mathcal{R}} r_t^{(k)}(z_t,\theta) z^{\chi(k)}_tdt\big) + o(dt)\nonumber
\end{IEEEeqnarray}
\end{center}

Under some regularity conditions detailed in \cite{Ethier1986}, \cite{Fuchs2013} or \cite{Guy2013}, and due to the density-dependance of transition rates, the dynamic of the system converges to a deterministic behaviour as the population size tends to infinity. For finite populations, the additional stochastic behaviour is termed \emph{demographic} stochasticity.

\subsection{Environmental stochasticity}

Extensions of the compartmental modeling framework introduced in the previous section have been proposed by the authors of \cite{Breto2009}, to account for additional sources of uncertainty related to fluctuations in extrinsic determinants of the epidemic that are not explicitly included in the model. This uncertainty is reflected through additional sources of stochasticity, termed \emph{environmental} stochasticity. The approach suggested in \cite{Breto2009} is to consider stochastic transition rates $\tilde{r}^{(k)}_t$ for a subset $\mathcal{R}^e$, under the following constraints for all $t$:

\begin{center}
\begin{IEEEeqnarray}{rCl}
\mathbb{E}\big(\tilde{r}^{(k)}_t\big) &\sim& r^{(k)}_t\nonumber\\
\tilde{r}^{(k)}_t &\geq& 0\nonumber
\end{IEEEeqnarray}
\end{center}

The presence of white noise in a model with stochasticitay $\sigma$ will be denoted $w.n.(\sigma)$ in the rows corresponding to noisy reactions. Yet, uncertain variations of extrinsic factors cannot always be modeled through high-frequency independent fluctuations. The evolution of climate, for example, has been shown to exhibit complex seasonal and inter-annual variations that influence epidemic dynamics \citep{Viboud2004}. Following the work of \cite{Cazelles1997} and \cite{Cori2009}, the authors of \cite{Dureau2013a} have proposed a general inferential framework for time-varying parameters, that is extended in the present document. Under this approach, parameters are modeled through stochastic differential equations or extensions thereof. The state vector is extended with additional components $x^{\theta_t}_t$ which dynamic is determined by the following equation: 
 
\begin{IEEEeqnarray}{rCl}
dx^{\theta_t}_t = \mu^{\theta_t}(x^{\theta_t}_t,\theta)dt + L^{\theta_t}dB_t^{Q^{\theta_t}}
\end{IEEEeqnarray}

Note that some constraints as positivity or boundedness generally need to be preserved when allowing parameters to vary over time, which is achieved by defining $x^{\theta_t}$  respectively as the log or logit transformation of the quantity of interest.

\subsection{Examples}

We introduce three models that will be used in the last section in different contexts (plague, H1N1 and dengue). These models are defined following the formalism that has just been defined, from which different mathematical formulations can be derived as will be described in the following Section.

\subsubsection{\label{sec:plague}Plague: SI model with seasonal forcing}

In this model, we consider two compartments: individuals are either infected or susceptible to be infected. Historical records show that plague is a seasonal forcing, consequently the reproduction rate is described as a periodic, sinusoidal function. If the life expectancy with plague is denoted with $\mu_D^{-1}$, the model can be described in the following way:

\begin{center}
\begin{tabular}{ccc}
\textbf{Reaction} & \textbf{Effect}   & \textbf{Rate}   \tabularnewline
\hline
infections  & $(S_t,I_t)\rightarrow (\boldsymbol{S_t-1},\boldsymbol{I_t+1})$ & $R_0[1+e\sin(t+\phi)]\mu_DI/N$  \tabularnewline
deaths       & $(S_t,I_t)\rightarrow (S_t,\boldsymbol{I_t-1})$ & $\mu_D$  \tabularnewline
\end{tabular}
\end{center}

\subsubsection{H1N1: SEIR model with time-varying contact rate}

We have illustrated in \cite{Dureau2013a} how it is possible to capture the evolution of key parameters of an epidemic, on the specific example of the 2009 H1N1 epidemic in London. The latter exhibits two waves, which is an unusual trajectory and suggests that the drivers of the epidemic have changed over time. As a first approach to this problem, an SEIR model can be used to reflect the fact that after being infected, individuals spend some time in a latent state before developing symptoms and becoming infectious. At the end of the infectivity period, recovered individuals become resistent and can no longer become infected:

\begin{center}
\begin{tabular}{ccc}
\textbf{Reaction} & \textbf{Effect}   & \textbf{Rate}   \tabularnewline
\hline
infections  & $(S_t,E_t,I_t,R_t)\rightarrow (\boldsymbol{S_t-1},\boldsymbol{E_t+1},I_t,R_t)$ & $\beta_t I/N$  \tabularnewline
onset of symptoms       & $(S_t,E_t,I_t,R_t)\rightarrow (S_t,\boldsymbol{E_t-1},\boldsymbol{I_t+1},R_t)$ & $k$  \tabularnewline
recovery       & $(S_t,E_t,I_t,R_t)\rightarrow (S_t,E_t,\boldsymbol{I_t-1},\boldsymbol{R_t+1})$ & $\gamma$  \tabularnewline
\end{tabular}
\end{center}

In order to capture the unknown variations of the effective contact rate $\beta_t$, it can be modelled using a random walk in the log space:

\begin{IEEEeqnarray}{rCl}
d\log \beta_t = \sigma dB_t
\end{IEEEeqnarray}

Note that in this example, transmissibility is directly reflected by the effective contact rate $\beta_t$, while a different parameterisation based on the reproduction rate $R_0$ and the infectivity period $\mu_D$ was used in the plague example. Both of these equivalent approaches can be found in the literature. 

\subsubsection{\label{sec:dengue}Dengue: parsmonious 2-strains model}

Dengue is a seasonal disease with complex dynamics, for which four strains co-exist. It is believed that after recovering from infection, individuals go through a short period of cross-immunity that protects them from all strains. Typical dengue case records do not specify with which strain individuals have been infected, hence a parsimonious two-strains model has been proposed in \cite{Aguiar2011} to temper identifiability issues:

\begin{center}
 \begin{adjustwidth}{-0.35in}{-0.35in}
\begin{tabular}{ccc}
\textbf{Reaction} & \textbf{Effect}   & \textbf{Rate}  \tabularnewline
\hline
1st infections  & $(S_t,I1,I2,R1,R2,S1,S2,I12,I21,R)$ & $\beta_t(I1/N+i+\psi I21/N)$  \tabularnewline
with str. 1&$\rightarrow (\boldsymbol{S_t-1},\boldsymbol{I1+1},I2,R1,R2,S1,S2,I12,I21,R)$ &  $w.n.(\sigma)$ \tabularnewline[0.5cm]
1st infections  & $(S_t,I1,I2,R1,R2,S1,S2,I12,I21,R)$ & $\beta_t(I2/N+i+\psi I12/N)$  \tabularnewline
with str. 2&$\rightarrow (\boldsymbol{S_t-1},I1,\boldsymbol{I2+1},R1,R2,S1,S2,I12,I21,R)$ &  $w.n.(\sigma)$ \tabularnewline[0.5cm]
recovery & $(S_t,I1,I2,R1,R2,S1,S2,I12,I21,R)$ & $\gamma$  \tabularnewline
from str. 1&$\rightarrow (S_t,\boldsymbol{I1-1},I2,\boldsymbol{R1+1},R2,S1,S2,I12,I21,R)$ &  \tabularnewline[0.5cm]
recovery & $(S_t,I1,I2,R1,R2,S1,S2,I12,I21,R)$ & $\gamma$  \tabularnewline
from str. 2&$\rightarrow (S_t,I1,\boldsymbol{I2-1},R1,\boldsymbol{R2+1},S1,S2,I12,I21,R)$ &  \tabularnewline[0.5cm]
loss of & $(S_t,I1,I2,R1,R2,S1,S2,I12,I21,R)$ & $\alpha$  \tabularnewline
cross-immunity&$\rightarrow (S_t,I1,I2,\boldsymbol{R1-1},R2,\boldsymbol{S1+1},S2,I12,I21,R)$ &  \tabularnewline[0.5cm]
loss of  & $(S_t,I1,I2,R1,R2,S1,S2,I12,I21,R)$ & $\alpha$  \tabularnewline
cross-immunity&$\rightarrow (S_t,I1,I2,R1,\boldsymbol{R2-1},S1,\boldsymbol{S2+1},I12,I21,R)$ &  \tabularnewline[0.5cm]
2nd infections  & $(S_t,I1,I2,R1,R2,S1,S2,I12,I21,R)$ & $\beta_t(I2/N+i+\psi I12/N)$  \tabularnewline
with str. 2&$\rightarrow (S_t,I1,I2,R1,R2,\boldsymbol{S1-1},S2,\boldsymbol{I12+1},I21,R)$ & $w.n.(\sigma)$ \tabularnewline[0.5cm]
2nd infections  & $(S_t,I1,I2,R1,R2,S1,S2,I12,I21,R)$ & $\beta_t(I1/N+i+\psi I21/N)$  \tabularnewline
with str. 1&$\rightarrow (S_t,I1,I2,R1,R2,S1,\boldsymbol{S2-1},I12,\boldsymbol{I21+1},R)$ & $w.n.(\sigma)$  \tabularnewline[0.5cm]
\end{tabular}
\end{adjustwidth}
\end{center}

 As with plague, seasonality is enforced explicitly through a sinusoidal factor: 
 $$\beta_t = \beta \times[1+e\sin(t+\phi)]$$ 
 This model allows for a different infectivity for individuals that have been infected for the second time, through the factor $\psi$. In addition, correlated \emph{white} environmental noise is enforced on infection reactions as a mean to reflect potential mis-specifications of the model.

\subsection{Tractable approximations of compartmental models}
\subsubsection{\label{subsub:ode}Ordinary differential equations}

The simplest and most stringent approximation of compartmental models are ordinary differential equations (ode's):

\begin{align}
	\frac{dz_t}{dt}&=  \sum_{k \in \mathcal{R}} l^{(k)}  r^{(k)} (z_t,\theta) z^{\chi(k)}_t
\end{align}

Under this formalism, the number of individuals in each compartment takes continues values, and varies continuously (and in a differentiable manner) over time. More specifically, all kind of demographic or environmental stochasticity are neglected, leading  the state of the system to evolve deterministically. From a practical perspective, the use of ordinary differential equations drastically simplifies the process of Bayesian inference, mainly due to the deterministic one-to-one mapping between trajectories $z_{0:T}$ and parameters $\theta$.

This formalism can be legitimately used for large populations and when all significant intrinsic and environmental factors have been explicitly incorporated in the deterministic skeleton of the model. However, in alternative cases results should be treated with caution, and the use of other formalisms accounting for  demographic or environmental stochasticity may be required.

\subsubsection{\label{subsub:sde}Stochastic differential equations}

Stochastic differential equations (sde's) are a natural extension of ode's, wherein state variables still take continuous values over time, and evolve continuously over time. Yet, trajectories of the system are no longer deterministic and differentiable due to the introduction of a driving Brownian motion reflecting the stochasticity of the system. To introduce this formalism, we rely on the notations used by the author of \cite{Sarkka2006} that will be helpful to handle and represent different and independent sources of stochasticity:

\begin{IEEEeqnarray}{rCl}
\label{eq:sde}
dx_t = \mu_t(x_t,\theta)dt + LdB_t^{Q_t}
\end{IEEEeqnarray}

In this equation, $\mu_t$ is referred to as the drift, $L$ as the dispersion matrix, and $Q_t$ as the diffusion matrix of the driving Browian motion. In particular, in our models the state variable $x_t$ is built from the concatenation of $z_t$ and $x^{\theta_t}_t$ respectively corresponding to the variables describing the structure of the population and to the variables monitoring the evolution of diffusing parameters over time. We can reformulate Eq. \ref{eq:sde}, utilising the notations that have been introduced earlier in this document:

\begin{IEEEeqnarray}{rCl}
dz_t &=& \sum_{k \in \mathcal{R}} l^{(k)}  r^{(k)} (z_t,\theta) z^{\chi(k)}_t dt + LdB_t^{Q_t}\nonumber\\
dx^{\theta_t}_t &=& \mu^{\theta_t}(x^{\theta_t}_t,\theta)dt + L^{\theta_t}dB_t^{Q^{\theta_t}}\nonumber
\end{IEEEeqnarray}

Note that the deterministic skeleton of the population variables' dynamic correspond to the ode model introduced in the previous section. The dispersion matrix $Q_t$ is a square matrix of size $n_{Q_t}\times n_{Q_t}$, and $L$ is a rectangular matrix of size $c\times n_{Q_t}$ ($c$ is the number of compartments in the model). Let us illustrate the use of these objects by introducing how demographic stochasticity can be incorporated in the model, based on the diffusion approximation, and further how the white noise environmental stochasticity  can be reflected.

\paragraph{Diffusion approximation of the demographic stochasticity}\mbox{}\\

In order to provide an SDE approximation of the demographic stochasticity, we rely on theoretical results of state-dependent Markov jump processes  presented in \cite{Ethier1986}. The adaptation of these results to compartmental epidemic models models has been illustrated in \cite{Fuchs2013}. Extensions of these results in non-homogeneous settings are provided in \cite{Guy2013}.

The diffusion approximation builds up on the definition of jump process models through their master equation:
\begin{IEEEeqnarray}{rCl}
\frac{\partial}{\partial t}P(z_t)=\sum_{k\in\mathcal{R}} r^{(k)} \tilde{z}_{k,t}^{\chi(k)}P(z_t-l^{(k)}) - \sum_{k \in\mathcal{R}} r^{(k)}(z_t,\theta) z^{\chi(k)}_t P(z_t) 
\end{IEEEeqnarray}

Where $\tilde{z}_{k,t}=z_t-l^{(k)}$.
The first term corresponds to the probability for the state vector of evolving into $z_t$, and the second corresponds to the probability of leaving the state $z_t$. In a SIR model setting, the master equation becomes:
\begin{IEEEeqnarray}{rCl}
\frac{\partial}{\partial t}P(S_t,I_t,R_t)&=& \beta \frac{(S_t+1)}{N}(I_t-1)P(S_t+1,I_t-1,R_t)  \nonumber \\
 & &  +\; \gamma (I_t+1)P(S_t,I_t+1,R_t-1)\\
 & &  - \; \beta \frac{S_t}{N}I_tP(S_t,I_t,R_t) \nonumber \\
 & &  - \; \gamma I_tP(S_t,I_t,R_t)\nonumber
\end{IEEEeqnarray}

This equation can be written in terms of normalised quantities, with $\varepsilon=1/N$:
\begin{IEEEeqnarray}{rCl}
\frac{\partial}{\partial t}P(s_t,i_t,r_t)&=& \frac{1}{\varepsilon} \beta (s_t+\varepsilon)(i_t-\varepsilon)P(s_t+\varepsilon,i_t-\varepsilon,,r_t)  \nonumber \\
 & &  +\;   \frac{1}{\varepsilon} \gamma (i_t+\varepsilon)P(s_t,i_t+\varepsilon,r_t-\varepsilon)\\
 & &  -\;   \frac{1}{\varepsilon} \beta s_t i_tP(s_t,i_t,r_t)  \nonumber\\
 & &  -\;   \frac{1}{\varepsilon} \gamma i_tP(s_t,i_t,r_t) \nonumber
\end{IEEEeqnarray}

The diffusion approximation relies on the limit of this expression when $\varepsilon \rightarrow 0$ while $N$ is kept constant. The author of \cite{Fuchs2013} shows that in this case, the former master equation converges to the following partial differential equation:
\begin{IEEEeqnarray}{rCl}
\frac{\partial}{\partial t}P(s_t,i_t,r_t)&=& \frac{\partial}{\partial s} \beta s_t i_t P(s_t,i_t,r_t)  -  \frac{\partial}{\partial i} (\beta s_t i_t -\gamma i_t)P(s_t,i_t,r_t)  \nonumber \\
 & &  +\;  \frac{1}{2} \frac{\partial^2}{\partial s^2} \frac{1}{N} \beta s_t i_t P(s_t,i_t,r_t)  \\
 & &  -\;  \frac{1}{2} \frac{\partial^2}{\partial i^2} \frac{1}{N} (\beta s_t i_t -\gamma i_t) P(s_t,i_t,r_t)   \nonumber\\
 & &  -\;   \frac{\partial^2}{\partial s \partial i} \frac{1}{N} \beta s_t i_t P(s_t,i_t,r_t)\nonumber,\label{eq:Ch2_pde}
\end{IEEEeqnarray}

which is equivalent to
\begin{IEEEeqnarray}{rCl}
\frac{\partial}{\partial t}P(s_t,i_t,r_t)&=& -\frac{\partial}{\partial x}[\dot{A}(s_t,i_t,r_t)P(s_t,i_t,r_t)] +  \frac{1}{2} \frac{\partial}{\partial x} \frac{\partial}{\partial x} [\dot{\Sigma}(s_t,i_t,r_t)P(s_t,i_t,r_t)]\label{eq:Ch2_pdegen}
\end{IEEEeqnarray}

\vskip0.5cm

Where
\begin{IEEEeqnarray}{rCl}
\;\;\dot{A}(s_t,i_t,r_t) = \left(\begin{array}{c} -\beta s_t i_t \\ \beta s_t i_t - \gamma i_t \\   \gamma i_t \end{array}\right)\end{IEEEeqnarray}

and

\begin{IEEEeqnarray}{rCl}
\dot{\Sigma}(s_t,i_t,r_t) = \frac{1}{N} \left(\begin{array}{ccc} \beta s_t i_t & -\beta s_t i_t & 0 \\ - \beta s_t i_t& \beta s_t i_t + \gamma i_t & - \gamma i_t   \\ 0 &  - \gamma i_t &  \gamma i_t  \end{array}\right)
\end{IEEEeqnarray}

\vskip0.5cm
Following  \cite{Kloeden1999},  Eq.  \ref{eq:Ch2_pdegen}  is a Fokker-Planck equation corresponding to a diffusion process that is a solution of
\begin{IEEEeqnarray}{rCl}
d\dot{z}_t = \dot{A}(\dot{z}_t)dt + LdB_t^{\dot{Q}_t^{d}} \label{eq:Ch2_sdenorm} 
\end{IEEEeqnarray}

Here, we follow the formalism of \cite{Sarkka2006}  where   $dB_t^{\dot{Q}_t^{d}}$ is a Brownian motion   with  diffusion matrix $\dot{Q}_t^{d}$ and $L$ is a stoichiometric dispersion matrix such that $L\dot{Q}_t^{d} L=\dot{\Sigma}$:  
\begin{IEEEeqnarray}{rCl}
\dot{Q}^d(s_t,i_t) = \frac{1}{N} \left(\begin{array}{cc} \beta s_t i_t & 0\\  0 & \gamma i_t  \end{array}\right)\;\;\;\; and \;\;\;\; L =  \left(\begin{array}{cc} -1  & \;\;\;0\\  \;\;\;1 & - 1 \\ \;\;\;0 & \;\;\;1 \end{array}\right)
\end{IEEEeqnarray}

Equation  \ref{eq:Ch2_sdenorm}  can be transposed in the natural scale of $z_t=[S_t,I_t,R_t]^T$, with $A=N\dot{A}$ and $Q^d = N^2\dot{Q^d} $:
\begin{IEEEeqnarray}{rCl}
dz_t = A(z_t)dt + LdB_t^{Q^d}
\label{eq:Ch2_SDE}
\end{IEEEeqnarray}

This result can be generalised based on the density-dependance property of rates $(r^{(k)}z_t^{\chi(k)})_{1\leq k \leq n}$. Formal proofs for  the general case of density-dependent jump processes can be found in \cite{Ethier1986}. The authors demonstrate that the dynamic of a density-dependent Markov jump process can be approximated with Eq.  \ref{eq:Ch2_SDE} with $dB$t being a multivariate Brownian motion with diffusion matrix $\dot{Q}^d = diag\{ r^{(k)}z_t^{\chi(k)},$ $k\in \mathcal{R}\}$, and L being the rectangular stoichiometric matrix which columns are the stoichiometric vectors $l^{(k)}$ with $k\in \mathcal{R}$.  Additionally, the drift component $\dot{A}(t)$ is determined by:
\begin{IEEEeqnarray}{rCl}
\dot{A}(z_t) = \sum_{k\in \mathcal{R}} l^{(k)}  r^{(k)}(z_t,\theta)\dot{z}_t^{\chi(k)}
\end{IEEEeqnarray}

Lastly, the resulting general  expression for $\dot{\Sigma}$ is the following:
\begin{IEEEeqnarray}{rCl}
\dot{\Sigma(z_t)} = L\dot{Q}^d L'=\sum_{k\in \mathcal{R}} l^{(k)}  r^{(k)}(z_t,\theta)\dot{z}_t^{\chi(k)} l^{(k)\prime}
\end{IEEEeqnarray}

\paragraph{Diffusion approximation of the environmental stochasticty}\mbox{}\\

This section focuses on environmental stochasticity. In this perspective, we consider for the sake of illustration an infinite population leading to a deterministic behaviour in the absence of diffusing parameters:
\begin{IEEEeqnarray}{rCl}
d z_t =   \sum_{k \in \mathcal{R}} l^{(k)}  r^{(k)} (z_t,\theta)z_t^{\chi(k)}dt
\end{IEEEeqnarray}

In the case of the SIR model:
\begin{equation}
\left\{
\begin{IEEEeqnarraybox}[
\IEEEeqnarraystrutmode
\IEEEeqnarraystrutsizeadd{6pt}
{2pt}
][c]{rCl}
dS_t &= &-\beta S_t \frac{I_t}{N} dt \\
dI_t &=& (\beta S_t \frac{I_t}{N}-\gamma I_t) dt\\
dR_t &=& \gamma I_t dt
\end{IEEEeqnarraybox}
\right.
\end{equation}

The framework proposed in \cite{Breto2009} introduces environmental stochasticity by replacing deterministic time increments $dt$ by random, stationary and nonnegative increments $d\Gamma_t$ with mean $dt$ and variance $\sigma^2 dt$. Here, if environmental noise is put over the infection reaction:
\begin{equation}
\left\{
\begin{IEEEeqnarraybox}[
\IEEEeqnarraystrutmode
\IEEEeqnarraystrutsizeadd{6pt}
{2pt}
][c]{rCl}
dS_t &= &-\beta S_t  \frac{I_t}{N} d\Gamma_t \\
dI_t &=& \beta S_t \frac{I_t}{N} d\Gamma_t -\gamma I_t dt \\
dR_t &=& \gamma I_t dt 
\end{IEEEeqnarraybox}
\right.
\end{equation}

We propose to derive a Gaussian formulation of epidemic models with \emph{white} environmental stochasticity by approximating $d\Gamma_t$ as $dt + \sigma dB_t$, i.e. the Gamma-distributed increments are replaced with a deterministic drift and a Brownian motion term with corresponding mean and variance. Thus, the model can be written as a stochastic differential equation:
\begin{equation}
\left\{
\begin{IEEEeqnarraybox}[
\IEEEeqnarraystrutmode
\IEEEeqnarraystrutsizeadd{6pt}
{2pt}
][c]{rCl}
dS_t &= &-\beta S_t \frac{I_t}{N} dt  - \sigma \beta S_t \frac{I_t}{N} dB^{(1)}_t \\
dI_t &=& (\beta S_t \frac{I_t}{N} d\Gamma_t -\gamma I_t)dt  + \sigma \beta S_t \frac{I_t}{N} dB^{(1)}_t\\
dR_t &=& \gamma I_t dt 
\end{IEEEeqnarraybox}
\right.
\end{equation}
\vskip0.5cm

In the general case, independent environmental noise can be enforced upon any subset $\mathcal{R}^e\in \mathcal{R}$ of all reactions:
\begin{IEEEeqnarray}{rCl}
dz_t =   \sum_{k \in \mathcal{R}} l^{(k)}  r^{(k)} (z_t,\theta)z_t^{\chi(k)}dt + L^edB_t^{Q^e}
\end{IEEEeqnarray}

$L^e$ is the $c\times Card(\mathcal{R}^e)$ stoichiometric matrix which columns are the stoichiometric vectors $l^{(k)}$ with $k\in \mathcal{R}^e$. In addition,  if all white noises are independent $dB_t^{Q^e}$ is a Brownian motion with diffusion matrix $Q^e= diag\big\{ \big(\sigma^{(k)}r^{(k)}(z_t,\theta)z_t^{\chi(k)}\big)^2,k\in \mathcal{R}^e\big\}$   containing the variance of the different environmental noises imposed upon the system.

In addition, it may be useful to enforce correlation between white noises affecting different reactions. In a multi-strain epidemic model, for example, if white noise is meant to capture climatic variability its impact may be the same on all transmission reactions. The latter can be achieved by the introduction of a second level of hierarchy accounting for grouping among noisy reactions. The latter can be determined through a mapping function $\varphi:\mathcal{R}^e\rightarrow[1:n_g]$  so that $\varphi^{-1}(p)$ corresponds to the indexes of  a group of correlated reactions for each $p\in [1:n_g]$. More details can be found in the following paragraph.

\paragraph{Diffusion approximation of compartmental models in the general case}\mbox{}\\

From the previous results, a diffusion approximation of compartmental models in the general case is provided by the following SDE:
\begin{equation}
\left\{
\begin{IEEEeqnarraybox}[
\IEEEeqnarraystrutmode
\IEEEeqnarraystrutsizeadd{6pt}
{2pt}
][c]{rCl}
dz_t &=&  \sum_{k \in \mathcal{R}} l^{(k)}  r^{(k)}(z_t,\theta)z_t^{\chi(k)}dt + LdB^{Q}_t\\
dx^{\theta_t}_t &=& \mu^{\theta_t}(x^{\theta_t}_t,\theta)dt + L^{\theta_t} dB^{Q^{\theta_t}}_t
\end{IEEEeqnarraybox}
\right.
\end{equation}

The matrices $L$ and $Q$ are constructed by concatenating the dispersion and diffusion matrices of the different sources of independent noises:
\begin{IEEEeqnarray}{rCl}
L = \left(\begin{array}{ccc}  L^d & L^e \end{array}\right) \;\;\;\; and \;\;\;\;  Q = \left(\begin{array}{ccc}  Q^d & 0 \\ 0 & Q^e \end{array}\right)
\end{IEEEeqnarray}

On one hand, $Q^d = diag\{ r^{(k)}(z_t,\theta)z_t^{\chi(k)},$ $k\in \mathcal{R}\}$ and $L^d = [l^{(1)},..,l^{(c)}]$ accounts for demographic stochasticity. With regards to \emph{white} environmental noise, $L^e=[l^{(k)}]_{k\in\mathcal{R}^e}$ is the concatenation of the stoichiometic vectors for noisy reactions. $\varphi$ is the mapping function defined over $\mathcal{R}^e$ that attributes an equal index in $[1;n_g]$ to reactions upon which correlated environmental noise is enforced. From this function, a rectangular $card(\mathcal{R}^e)\times n_g$ dispersion matrix $L^g$ can be constructed in which the column of group $p$ is filled with $r(z_t,\theta)z_t^{\chi(k)}$ on rows corresponding to reactions such that $\varphi(k)=p$, and zero's everywhere else. With $Q^g=diag\big\{ \big(\sigma^{(p)})^2,p\in [1:n_g]\big\}$, $Q^e$ can be computed as $Q^e=L^g Q^g L^{g\prime}$. Naturally, this method for constructing correlated noise terms hold for uncorrelated noises.

\subsubsection{\label{subsub:psr}Poisson process with stochastic rates}

The continuous approximation of the number of individuals contained in each compartment, and of its evolution, may be questionable when populations at stake are not large enough and more specifically when the size of at least one compartment becomes small. Such situations typically correspond to the extinction of diseases or species in epidemic or ecological models. The Markovian jump process introduced earlier accounts for the discrete nature of the size of each compartment, and the discontinuities induced by the occurrence of each reaction. Nevertheless, due to the density-dependence of transformation rates the frequency of reactions increases infinitely as $N\rightarrow\infty$. Hence, the reference Markov jump process formalism quickly becomes intractable for other than small populations. The authors of \cite{Breto2009} have proposed an approximation of the Markov jump process based on a multinomial approximation of the number of reactions occurring over a short period of time $dt$. Here, we reformulate the solution proposed in \cite{Breto2009} and extend it to the general framework for compartmental models proposed in SSM.

The Poisson process model determines the probability that each reaction $k$ ($k\in\mathcal{R}$) respectively occurred $n_k$ times over a given period $dt$. If all sources of environmental stochasticity are neglected:

\begin{IEEEeqnarray}{rCl}
p(n_1,\dots,n_m|z_t,\theta) = \prod_{i=1}^c\left\{  M_{i}  \left( 1-\sum_{\chi(k)=i} p_{k} \right) ^{\overline{n}_{i}}  \prod_{\chi(k)=i} \left(  p_{k} \right)^{n_{k}}   \right\} + o(dt)\nonumber
\end{IEEEeqnarray}

Using the following notations: 
\begin{IEEEeqnarray}{rCl}
&&p_{k}=p_{k}  \left(  r^{(k)}(z_t,\theta)z^{\chi(k)}_t dt\right) = \left( 1-\exp \left\{-\sum_{\chi(k')=i}r^{(k')}(z_t,\theta)z^{\chi(k')}_t dt \right\} \right)\frac{r^{(k)}(z_t,\theta)}{\sum_{\chi(k')=i} r^{(k')}(z_t,\theta)}\nonumber\\
&&\overline{n}_{i}=z_t^{(i)} - \sum_{\chi(k)=i} n_{k}\nonumber\\
&&M_{i} = \left( \begin{array}{c} z_t^{(i)} \\ \{n_{k}\}_{\chi(k)=i} \; \overline{n}_i  \end{array}  \right) \;\;\;\;\;\text{
(multinomial coefficient)}\nonumber
\end{IEEEeqnarray}

In addition, white noise can be introduced on reaction $k$ ($k\in\mathcal{R}^e$) by replacing time increments $dt$ by random increments $d\Gamma_k$ with Gamma distribution, mean $dt$ and standard deviation $\sigma^{(k)}\sqrt{dt}$:

\begin{IEEEeqnarray}{rCl}
&&p_{k}=p_{k}  \left(  r^{(k)}z^{\chi(k)}_t d\Gamma_k \right) = \left( 1-\exp \left\{-\sum_{\chi(k')=i}r^{(k')}(z_t,\theta)z^{\chi(k')}_t d\Gamma_{k'} \right\} \right)\frac{r^{(k)}d\Gamma_k}{\sum_{\chi(k')=i} r^{(k')}d\Gamma_{k'}}\nonumber
\end{IEEEeqnarray} 

Lastly, time-varying parameters can be introduced in a similar manner as under previous formalisms:

\begin{IEEEeqnarray}{rCl}
dx^{\theta_t}_t = \mu^{\theta_t}(x^{\theta_t}_t,\theta)dt + L^{\theta_t}dB_t^{Q^{\theta_t}}\nonumber
\end{IEEEeqnarray}

\section{\label{sec:methods}Library of inference methods}

In this section we will consider the more general class of state space models evolving in continuous time, with discrete observations. This definition encompasses systems of ordinary or stochastic differential equations as well as compartmental models and combinations thereof. The state of the system at time $t$ is noted $x_t$. We will abusively note $x_i$ the value of $x_t$ at time $t_i$, hence $x_{0:n}$ denotes a trajectory of the system between $t_0$ and $t_n$. The prediction density $p(x_{i+1}|x_i,\theta)$ is generally untractable, which means that the probability of getting to state $x_{i+1}$ from state $x_{i}$ cannot be computed. However, we consider that it is possible to simulate trajectories from the augmented prediction density $p(x_{i:i+1}|x_i,\theta)$.
In addition, an observation model $p(y_i|x_i,\theta)=f(h(x_i);y_{i},\theta)$ needs to be defined, to determine what is the probability of observing $y_i$ conditionnally on the set of parameters $\theta$ and its proxy $h(x_i)$ built from the state of the system $x_i$.

\subsection{Inference for state space models}

State space models can be seen as a hypothesised probabilistic relation between the trajectories $x_{0:n}$ of a system and constant related  quantities grouped in a parameter vector $\theta$. This relation determines a joint probability density $p(x_{0:n},\theta)$. From a Bayesian perspective, the knowledge or the uncertainty over the components of $\theta$ are enforced through the \emph{a priori} density $p(\theta)$. For a given parameter vector $\theta$, the likely trajectories of the system are reflected by the density $p(x_{0:n}|\theta)$. The primary objective of inference with state space models is the estimation of the posterior density $p(\theta|y_{1:n})$, and of the marginal density $p(x_{0:n}|y_{1:n})$. Additionally, model choice indicators can play a key role in disentangling between different hypothesised state space models \citep{Spiegelhalter2002}.

As suggested by the now classic motto "all models are wrong, but some are useful" \citep{Box1987}, models will only ever be a rough approximation of a complex reality. Yet, the latter does not prevent from following a scientific inductive approach to derive conclusions from the confrontations of models to data. By reconstructing the trajectory $x_{0:n}$ of the partially observed system or learning about uncertain components of $\theta$, experience suggests that this process is likely to revise our understanding of infectious diseases \citep{King2008}.  As in any other context, the validity of inference results shall be critically examined at least from a three-fold perspective. First, the uncertainties associated with the data collection should be reflected in the observation model. Then, the limitations of the model itself should be acknowledged and questioned, while considering the practical feasibility of  proposing extensions to palliate the imperfections of the model. A minimal condition requires the output of the model to be able to fit the available observations of mechanisms they are meant to reproduce \citep{Gelman2012}. At last, the information derived regarding $x_{0:n}$  and  $\theta$, reflected by the discrepancies between their marginal prior and posterior densities, should not be considered as hard truth but rather as plausible and testable hypothesis  \citep{Popper2002}.

An additional dimension arises when working with state space models, that requires specific attention. Although the joint posterior density $p(x_{0:n},\theta|y_{1:n})$ can  be computed  up to a multiplicative constant through the Bayes rule for a given trajectory $x_{0:n}$ and parameter $\theta$, there is generally no direct way of deriving tractable formulas for the quantities of interest, i.e. $p(\theta|y_{1:n})$ and $p(x_{0:n}|y_{1:n})$. For sufficiently small-dimensional problems, efficient solutions for routine inference are offered by Gibbs MCMC samplers as the ones implemented in the Bugs library \citep{Lunn2000}. In its current version, the \emph{SSM} library provides a tailored and more efficient  solution constructed around the particle Marginal Metropolis Hastings algorithm (pMMH), which is one of the two versions of the particle Markov Chain Monte Carlo algorithm (pMCMC) \citep{Andrieu2010}. We will introduce the different inference tools in the remainder of this Section, and motivate and illustrate their combination in the following one.

\subsection{Conditional state exploration: $p(x_{0:n}|y_{1:n},\theta)$ and $p(y_{1:n}|\theta)$}

\paragraph{Sequential Monte Carlo}\mbox{}\\
Sequential Monte Carlo (SMC) methods, also known as particle filters in this setting, provide an efficient solution to explore the space of trajectories of a system conditioned on parameters $\theta$ and available observations $y_{1:n}$. They are targeted to problems where the target density  can be decomposed as a product of terms. These terms are aggregated progressively in order to  achieve a smooth transfer from a simple initial density corresponding to a single term of the product, up to the full target density. For example, for state space models the algorithm starts by approximating the prior density of initial conditions $p(x_0|\theta)$ with a swarm of samples called particles. At each iteration of the algorithm, an additional observation is accounted for, progressively increasing the dimension of the explored state. Particles are weighted according to how well they fit the new datapoint, and a resampling step is made to ensure that the exploration focuses on informative regions of the target space.

A classic version of the SMC algorithm, referred to as Systematic Importance Resampling algorithm, is presented in Algorithm \ref{alg:SMC} \citep{Doucet2009}. If $J$ is the number of particles,  this algorithm can provide  a sample $\tilde{x}_{0:n}$ from $\hat{p}^J_{pf}(x_{0:n}|y_{1:n })$, and an unbiased estimator $\hat{p}^J_{pf}(y_{1:n}|\theta)$ of $p(y_{1:n}|\theta)$. Under mild assumptions, the authors of \cite{Moral2004} and \cite{Andrieu2010} have proved the following properties:
\begin{IEEEeqnarray}{rCl}
	\Vert \hat{p}^J_{pf}(x_{0:n}|y_{1:n }) - p(x_{0:n}|y_{1:n }) \Vert &\leq& \frac{C_n}{J}\\
	Var(\frac{\hat{p}^J_{pf}(y_{1:n}|\theta)}{p(y_{1:n}|\theta)}) &\leq&  \frac{D_n}{J}\nonumber
\end{IEEEeqnarray}

Where $C_n$ and $D_n$ are constants depending on the model and on the number of observations $n$. The distance $\Vert p_2-p_1 \Vert$  is defined as the total variation distance between the two distributions. Consequently, the particle filter is a solution to achieve asymptotically exact estimation of
the  marginal likelihood with precision increasing as $O(J^{1/2})$. 
\begin{algorithm}
\caption{Sequential Monte Carlo algorithm}
\label{alg:SMC}
{\fontsize{12}{20}\selectfont
\begin{algorithmic}
\STATE Set $L=1$, $W_{0}^{(j)}=\frac{1}{J}$, sample $(x_{0}^{(j)})_{j=1,...,J}$ from $p(x_0|\theta)$ 
\FOR {$k=0$ to $n-1$}
	\FOR {$j=1$ to $J$}
		\STATE Sample $(x_{k:k+1}^{(j)})$ from $p(x_{k:k+1}|x_k,\theta)$  
		\STATE Set $\alpha^{(j)}= h(y_{k+1},x^{(j)}_{k+1},\theta)$
	\ENDFOR
	\STATE Set $W_{k+1}^{(j)}=\frac{\alpha^{(j)}}{\sum_{l=1}^{J}\alpha^{l}}$, and $L=L\times \frac{1}{J} \sum_j \alpha^{(j)}$
	\STATE Resample $(x_{0:k+1}^{(j)})_{j=1,\dots,J}$ according to $(W_{k+1}^{(j)})$,
\ENDFOR		
\end{algorithmic}
}
\end{algorithm}

\paragraph{Extended Kalman Filter}\mbox{}\\
An approximate solution to the filtering problem for state space models is provided by the Extended Kalman Filter (EKF) algorithm \citep{Jazwinski1970,Sarkka2006}. We consider its continuous-discrete version tailored  to dynamic models formulated as stochastic differential equations, with $\mu$  corresponding to the drift component of the model (which Jacobian is noted $\nabla\mu$), and diffusion and dispersion matrices being respectively noted $Q$ and $L$. $R_k$ is the variance of the observation process at time $k$. The EKF, described in Algorithm \ref{alg:EKF},  is based on a gaussian approximation of the observation process $h$ (which Jacobian is noted $\nabla h$), resulting in a multivariate normal filtered density for $p(x_i|y_{0:i})$ characterised by its mean $m_t$ and covariance $C_t$. It provides with a deterministic and biased estimate $\hat{p}_{EKF}(y_{1:n}|\theta)$ of the marginal likelihood.

Note that in Algorithm \ref{alg:EKF},  only one observation is integrated at each time step. In the case of simultaneous observations, the same steps can be followed several time to update iteratively the mean and covariance of the state vector, observation per observation.

\begin{algorithm}
\caption{Continuous-discrete Extended Kalman Filter algorithm}
\label{alg:EKF}
{\fontsize{12}{20}\selectfont
\begin{algorithmic}
\STATE Set $L=1$ and initialise the mean state $m_t$ and covariance $C_t$
\FOR {$k=1$ to $n$}
	\STATE Integrate between $t_{k-1}$ and $t_{k}$:\\
		$\;\;\;\;\;\;\frac{dm_t}{dt}=\mu(m_t,\theta)$\\
		$\;\;\;\;\;\;\frac{dC_t}{dt}=\nabla\mu(m_t,\theta)C_t + C_t \nabla\mu(m_t,\theta)^T+LQL'$
	\STATE Compute the prediction error $e = y_{k}- h(m_{t_{k}},\theta)$, and the following quantities:\\
		$\;\;\;\;\;\;S=\nabla h(m_{t_{k}},\theta) C_{t_k} \nabla h(m_{t_{k}},\theta)'+R_{t_k}$	\\
		$\;\;\;\;\;\;K=C_{t_k}\nabla h'(m_t,\theta)S^{-1}$\\
	\STATE Update the mean state and Covariance:\\
		$\;\;\;\;\;\;m_t = m_t + K e$\\
		$\;\;\;\;\;\;C_t = C_t - KSK'$	
	\STATE Update the likelihood $L(\theta) = L(\theta)\times\mathcal{N}(e;0,S)$
\ENDFOR		
\end{algorithmic}
}
\end{algorithm}

\subsection{Full inference of paths and parameters}

The central methodology utilised in SSM to estimate the paths and parameters of compartmental models is the pMMH version of the pMCMC. For the sake of completeness, and for readers that may not be familiar with this methodology, we start by a brief introduction to the Monte Carlo Markov Chain machinery.

\subsubsection{Introduction to the Monte Carlo Markov Chain machinery}

Monte Carlo Markov Chain (MCMC) methods are used to estimate properties of probability densities in cases where analytic formulas cannot be directly derived, and samples cannot be directly generated. If we generically note $x$ ($x\in\mathbb{R}^d$) the  random variable of a target density  $\pi(.)$, MCMC algorithms only require the ability to compute $\pi(x)$ for any $x$, up to a multiplicative factor. Their founding mechanism is the construction of a Markov chain that randomly explores $\mathbb{R}^d$ taking values $(x^{(1)},x^{(2)},..,x^{(N)})$ which will asymptotically ($N\rightarrow\infty$) mimic samples drawn from the target distribution. The chain is defined through a transition kernel $K$ that determines the transition probability $p(.|x^{(i-1)})$. The chain converges to an invariant distribution if K is irreducible (from any state there is a positive probability to visit any other state) and aperiodic. The detailed balance condition is a sufficient but not necessary condition to ensure that the invariant distribution of the chain is the target density $\pi$:
\begin{align}
	\pi(x^i)K(x^{(i-1)}|x^{(i)}) = \pi(x^{(i-1)})K(x^{(i)}|x^{(i-1)})
\end{align}

A critical dimension of MCMC algorithms is their efficiency in $mixing$, i.e. in generating samples that are  as independent as possible. Unless $K(.|x^{(i)})$ is equal to  $\pi(.)$, $N$ samples of the MCMC trajectory will not provide the same amount of information as $N$ independent and identically distributed (i.i.d.) samples from the target density $\pi$. This can be quantified by the \emph{Effective Sample Size} (ESS), for example, that estimates how many truly i.i.d. samples the MCMC  output is equivalent to \citep{Geyer1992, Brooks1998}. Here is one way to compute the ESS:
\begin{align}
	ESS(\{x^{(1)},x^{(2)},..,x^{(N)}\}) = \frac{N}{1+2\sum_{k=1}^{k^{max}}Correl(\{x^{(1)},..,x^{(N-k)}\},\{x^{(k)},..,x^{(N)}\})}
\end{align}

This indicator, along with other diagnostic tools proposed in the CODA package \citep{Plummer2006}, are crucial in assessing the validity of results obtained through MCMC exploration of the complex and high-dimensional target density $p(x_{0:n},\theta|y_{0:n})$. Before diving into the presentation of basic and more advanced MCMC algorithms, we introduce the most classic way to define transition kernels that respect the detailed balance condition: the Metropolis-Hastings step \citep{Metropolis1953,Hastings1970}. At each iteration of the chain, a proposed value $x^*$ is sampled from an importance distribution $q(.|x^{(i)})$, and accepted with probability:
\begin{align}
	1\wedge \frac{\pi(x^*)q(x^{(i)}|x^*)}{\pi(x^{(i)})q(x^*|x^{(i)})}
\end{align}
Otherwise, $x^*$ is rejected and $x^{(i+1)}$ is set equal to $x^{(i)}$. The proportion of proposed samples that have been accepted determine the acceptance rate.  The Metropolis Hastings step allows the use of any importance distribution $q$ respecting the irreducibility and aperiodicity conditions, although other choices are also possible. It is generally observed that increasing the dimension of $x$ decreases the acceptance probability.

For example, the random walk Metropolis is  based  on a Metropolis-Hastings step using a multivariate normal importance sampling distribution: $q(.|x^{(i)})=\mathcal{N}(x^{(i)},\Sigma^q)$ (see Algoritm \ref{alg:Intro_MH}). The efficiency of this algorithm on a given problem depends on the calibration of the covariance matrix $\Sigma^q$.
Theoretical results have been demonstrated in the situation where the target  distribution $\pi$  is  a multivariate normal density:

\begin{prop}
	When $\pi$ is  a multivariate normal density, the acceptance rate that maximises the mixing efficiency of the random walk Metropolis algorithm is $23.4\%$ \citep{Roberts1997}
\end{prop}

\begin{prop}
	When $\pi$ is  a multivariate normal density, optimal results are achieved by using $\Sigma^q = \frac{2.38^2}{d}\times Cov(\pi)$ \citep{Roberts1997}.
\end{prop}

\begin{algorithm}[h]
\caption{random walk Metropolis algorithm}
\label{alg:Intro_MH}
{\fontsize{12}{20}\selectfont
\begin{algorithmic}
\STATE Initialise $x^{(0)}$
\FOR {$i=0$ to $N$}
	\STATE Sample $x^*\sim \mathcal{N}(x^{(i)}, \Sigma^q)$
	\STATE Accept $x^*$ with probability $1\wedge\frac{\pi(x^*)}{\pi(x^{(i)})}$
\ENDFOR
\end{algorithmic}
}
\end{algorithm}

When the target distribution is not a multivariate normal density, these results are generally extrapolated and followed as rules of conduct. They were used to derive adaptive versions of the random walk Metropolis algorithm, based on a decomposition of $\Sigma^q$ into $\lambda\Sigma$. A first adaptive algorithm exploits the monotonicity of the acceptance rate as a function of $\lambda$. The Metropolis-Hastings ratio of a random walk Metrpopolis algorithm is $\frac{\pi(x^*)}{\pi(x)}$. Hence, if the mass of the target density is concentrated in a certain region and $x$ is in this region (which is the case with high probability if the chain has converged), increasing the value of $\lambda$ increases the risk for $x^*$ to escape that region, leading to low values of $\pi(x^*)$ and rejection of $x^*$. On the contrary, excessively small values of $\lambda$ will induce values of $\frac{\pi(x^*)}{\pi(x)}$ close to one and high acceptance rates. Therefore, the targeted acceptance rate can be approached by iteratively adapting $\lambda$ with a cooling rate $a\in[0;1[$:
\begin{align}
	\lambda_{i+1} = \lambda_{i}\times a^i(AccRate_i-0.234)
\end{align}

A second adaptive algorithm relies on the fact that, as the chain progresses, the generated samples are meant to mimick i.i.d. samples generated from the target distribution $\pi$. Consequently, the empirical covariance matrix obtained from these samples can be used as a proxy for the optimal covariance matrix $\frac{2.38^2}{d}\times Cov(\pi)$. The resulting adaptive algorithm proposed in \cite{Roberts2009} is based on the following importance sampling distribution:
\begin{align}
q(.|x^{(i)})=\alpha \mathcal{N}\left(x^{(i)},\lambda\frac{2.38^2}{d}\Sigma^{(0)}\right) +(1-\alpha)\mathcal{N}\left(x^{(i)},\lambda\frac{2.38^2}{d}\Sigma^{(i)}\right),
\end{align}

with $\Sigma_i$ being the empirical  covariance matrix obtained from the $i$ samples generated by the chain. The use of a mixture of normal distributions ($\alpha$ is generally set to 0.05) is meant to avoid convergence to local modes.

\subsubsection{Particle Marginal Metropolis Hastings algorithm}
Sequential Monte Carlo techniques are a natural choice to explore $p(x_{0:n}|y_{1:n},\theta)$. In order to  account for  uncertainties regarding the parameter vector $\theta$, we are aiming for the exploration of the joint posterior density $p(x_{0:n},\theta|y_{1:n})$. 
The augmented path $x_{0:n}$, in particular, is a high-dimensional object. It contains the state of the system at each point of the discretised time $(t_0, t_{0}+\delta, t_{0}+2\delta,\dots,t_n-\delta,t_n)$. As previously mentioned,  classic MCMC methods fail to be efficient and robust solutions because of the high dimension of the target density. The particle MCMC algorithm  offers a solution relying on the efficiency of particle filters \citep{Andrieu2010}. Algorithm \ref{alg:PMCMC} illustrates the principles of its particle marginal Metropolis Hastings version: the high-dimensional density exploration problem is reduced to the design of an MCMC algorithm over $\theta$, based on the likelihood $\hat{p}^J_{pf}(y_{1:n}|\theta)$ estimated by a particle filter conditioned on $\theta$. The authors of \cite{Andrieu2010} have shown that for any $J$ the algorithm was asymptotically exact for a given discretisation of time. Under classic assumptions, when the number of iterations $N^{\theta}$ tends to infinity:
\begin{align}
	\Vert \hat{p}^J_{pf}(x^{(i)}_{0:n},\theta^{(i)}|y_{1:n}) -  p(x_{0:n},\theta|y_{1:n}) \Vert \rightarrow 0 \;\;\;\;\; as \; i\rightarrow \; \infty
\end{align}

Every iteration of the MCMC algorithm implies running a particle filter to explore the range of likely paths of the system under the current value of $\theta$ and observed data $y_{1:n}$. Consequently, the pMCMC is a computationally demanding algorithm; its complexity if of the order of $O(n J  N^{\theta})$. The mixing efficiency of the MCMC scheme critically determines the applicability of the algorithm. In the absence of suitable techniques to efficiently estimate the marginal score $\nabla_{\theta} \log p(\theta|y_{1:n})$, random walk Metropolis algorithms are generally used. Even in its adaptive form, the parameterisation of its initial covariance matrix $\Sigma^q_0$ is a central issue: we will explore in the next subsection a mean to automate this process, rendering the pMCMC algorithm plug-and-play.

\begin{algorithm}[h]
\caption{Particle MCMC algorithm (particle marginal Metropolis Hastings version)}
\label{alg:PMCMC}
{\fontsize{12}{20}\selectfont
\begin{algorithmic}
\STATE Initialise $\theta^{(0)}$.
\STATE Use the SMC algorithm to compute  $\hat{p}^J_{pf}(y_{1:n}|\theta^{(0)})$ and sample $x^{(0)}_{0:n}$ from $\hat{p}_{pf}^J(x_{0:n}|y_{1:n},\theta^{(0)})$
\FOR {$i=1$ to $N^{\theta}$}
	\STATE Sample $\theta^{*}$ from $q(.|\theta^{(i)})$
	\STATE Use the SMC to compute $L(\theta^*)=\hat{p}^J_{pf}(y_{1:n}|\theta^*)$ and sample $x_{0:n}^{*}$ from $\hat{p}_{pf}^J(x_{0:n}|y_{1:n},\theta^*) $
	\STATE Accept $\theta^{*}$ (and  $x_{0:n}^{*}$) with probability $1\wedge\frac{L(\theta^{*})q(\theta^{(i)}|\theta^{*})}{L(\theta^{(i)})q(\theta^{*}|\theta^{(i)})}$
	\STATE Record $\theta^{(i+1)}$ and $x^{(i+1)}_{1:n}$
\ENDFOR
\end{algorithmic}
}
\end{algorithm}

An alternative solution is the $SMC^2$ algorithm presented in \cite{Chopin2012}. It explores both the probability density of $x_{0:n}$ and $\theta$ with an SMC algorithm, starting from the initial target $p(x_0,\theta)$, and progressively incorporating  the available observations. The global complexity of this algorithm is similar to the pMCMC, but its ability to automatically adapt the number particles being utilised and to progressively learn from previous samples what could be seen as the equivalent of the covariance matrix $\Sigma^q$ are promising features. It additionally provides a estimate of $p(y_{1:n})$ under a given model, which can be used for model selection through the Bayes rule.

\subsubsection{Efficiently initialisation and calibration of the PMMH algorithm}

As we just mentioned, each iteration of the pMCMC algorithm is computationally demanding. For this reason, we want to reduce the calibration period of the pMCMC itself (also known as \emph{burn-in} period), which can be done by preliminary pre-explorations of the target density $p(\theta|y_{1:n})$. When MCMC chains are initialised from an arbitrary position, the likelihood classically follows an increasing trend  before it stabilises, indicating that the chain has converged to a mode. During this phase, $\theta$ also follows a transient convergence phase. Naturally, due to this non-stationarity the generated samples are strongly correlated and weakly informative. It is then natural to rely on optimisation algorithms to accelerate this transient phase and directly launch the pMCMC chain close from a mode of the posterior density. In addition, complex target densities generally exhibit local modes in which MCMC or optimisation algorithms can be trapped, again leading to false and misleading results. The search for a global mode is a challenging problem in itself and should be done with suitable and dedicated tools, that we will now present. In addition, the pMCMC implementation proposed in \emph{SSM} allows for the adaptation of the sampling covariance $\Sigma^q$. This adaptation phase can be long and costly, due to a "chicken and egg" situation: adaptation of $\Sigma^q$ is most needed when mixing is poor, which is also the situation where learning is the slowest. This issue will also be covered in this Section.

\paragraph{Searching for the global mode with the simplex, ksimplex and mif algorithms}\mbox{}\\
One of the most classic algorithms that can be used to optimise a function of continuous variables in an unconstrained space is the simplex algorithm, also known as the  Nelder-Mead algorithm (see Algorithm \ref{alg:simplex}). In \emph{SSM}, to avoid the complications that arise when introducing constraints (positivity or boundedness, for example) the components of $\theta$ are transformed through log or logit functions (or extensions thereof, to allow for boundaries different than 0 and 1) before being handed to the simplex algorithm. The latter operates by constructing a polygon with $d+1$ vertices, with $d=dim(\theta)$, and optimising the value of the target function at each of its vertices (in our case, $p(\theta|y)$) through reflection, expansion, contraction or reduction transformations of the polygon. The use of this algorithm in SSM directly relies on its implementation in the GNU Scientific Library \citep{Galassi2006}. The complexity of this algorithm increases linearly with d. In addition, it is a local exploration algorithm, and although it does not strictly follow the gradient of the target density, it can easily be trapped in local modes. At last, the simplex algorithm requires the target function $p(\theta|y)$ to be computed deterministically: it cannot be directly plugged to the particle filter where the estimation of the likelihood is noisy. As a consequence, the \texttt{simplex} algorithm can only be used in \emph{SSM} on ode approximations of the system (see \ref{subsub:ode}). In order to account for demographic or environmental sources of stochasticity, it is possible to estimate the likelihood with the Extended Kalman Filter from an sde approximation of the model (see \ref{subsub:sde}). As this estimate can be obtained deterministically, it can be plugged into the simplex algorithm: this is the \texttt{ksimplex} function available in  \emph{SSM}.

The optimisation routines based on algorithm \ref{alg:simplex} require the use of ode or sde approximations of the system. These may lead to biased estimated of the optimal set of parameters, with regards to what could be found using a psr formalism (see \ref{subsub:psr}). However, they are only used as a first step to initialise the Markov Chain that subsequently explores the posterior density:

\begin{align}
\texttt{cat theta.json | ./ksimplex | ./kmcmc | ./pmcmc}\nonumber
\end{align}

As a consequence, potential discrepancies between the likelihoods induced under the different formalisms will only have serious consequences in  critical cases. In such situations, it is possible to rely on iterated filtering (\texttt{mif}), an asymptotically exact and plug-and-play  solution to the frequentist problem of  maximising the marginal likelihood $p(y_{1:n}|\theta)$ \citep{Ionides2006,Breto2009,Ionides2011}. This approach has already been used for numerous applications in epidemiology \citep{Ionides2006,King2008,Breto2009,Laneri2010,He2010,Camacho2011,He2011}. Although it has been developed and utilised as a purely frequentist algorithm,  the mif  can also it can  be used to efficiently initialise the Markov chain of the PMCMC by incorporating the prior density into the maximised function of $\theta$ as illustrated in Algorithm \ref{alg:mif}.  Under the current implementation, corresponding to the algorithm described in \cite{Breto2009}, careful parameterisation of the algorithm is required to achieve convergence to the mode. Further investigation is being carried out to increase the efficiency and stability of the iterated filtering algorithm \citep{Ionides2012,Lindstrom2013}, which could allow to directly optimise the posterior distribution under the optimal psr formalism, exploiting the interesting \emph{tempering} feature of this approach. In the meantime, serialised \texttt{simplex} or \texttt{ksimplex} algorithms permit easier routine maximisation of the posterior density.

\begin{algorithm}[H]
\caption{Simplex algorithm (a.k.a. Nelder-Mead algorithm)}
\label{alg:simplex}
{\fontsize{12}{20}\selectfont
\begin{algorithmic}
\STATE Initialise $(\theta^{(1)},\dots,\theta^{(d+1)})$, with $d=dim(\theta)$.
\STATE Set $N=0$.
\STATE Unless stated otherwise, set $\alpha = 1,\gamma=2,\rho = -1/2$, and $\sigma=1/2$
\WHILE {convergence is not achieved and $N<N_{max}$}
	\STATE N = N+1
	\STATE Order according to the values at the vertices $p(\theta^{(d+1)}|y)\leq\dots\leq p(\theta^{(0)}|y)$
	\STATE Calculate $\theta^{(0)}$, the center of gravity of all points but $\theta^{(d+1)}$
	\STATE \textbf{Reflection}\\
	$\;\;\;\;\;\;$ Compute reflected point $\theta^{(r)}=\theta^{(0)}+\alpha(\theta^{(0)}-\theta^{(d+1)})$\\
	$\;\;\;\;\;\;$ If $p(\theta^{(d)}|y)\leq p(\theta^{(r)}|y)$ and $p(\theta^{(r)}|y)< p(\theta^{(0)}|y)$:\\
	$\;\;\;\;\;\;$$\;\;\;\;\;\;$ replace $\theta^{(d+1)}$ by $\theta^{(r)}$ and end iteration.
	\STATE \textbf{Expansion}\\
	$\;\;\;\;\;\;$ If $p(\theta^{(1)}|y)\leq p(\theta^{(r)}|y)$:\\
	$\;\;\;\;\;\;$$\;\;\;\;\;\;$ If $p(\theta^{(r)}|y)\leq p(\theta^{(e)}|y)$:\\
	$\;\;\;\;\;\;$$\;\;\;\;\;\;$$\;\;\;\;\;\;$ Compute the expanded point $\theta^{(e)}=\theta^{(0)}+\alpha(\theta^{(0)}-\theta^{(d+1)})$,\\
	$\;\;\;\;\;\;$$\;\;\;\;\;\;$$\;\;\;\;\;\;$ and replace $\theta^{(d+1)}$ by $\theta^{(e)}$, and end iteration.\\
	$\;\;\;\;\;\;$$\;\;\;\;\;\;$ Else:\\
	$\;\;\;\;\;\;$$\;\;\;\;\;\;$$\;\;\;\;\;\;$ Replace $\theta^{(d+1)}$ by $\theta^{(r)}$, and end iteration.
	\STATE \textbf{Contraction}\\
	$\;\;\;\;\;\;$ We know that $p(\theta^{(r)}|y)\leq p(\theta^{(d)}|y)$.\\
	$\;\;\;\;\;\;$ Compute the contracted point $\theta^{(c)}=\theta^{(0)}+\rho(\theta^{(0)}-\theta^{(d+1)})$,\\
	$\;\;\;\;\;\;$$\;\;\;\;\;\;$ If $p(\theta^{(d+1)}|y)\leq p(\theta^{(c)}|y)$:\\
	$\;\;\;\;\;\;$$\;\;\;\;\;\;$$\;\;\;\;\;\;$ Replace $\theta^{(d+1)}$ by $\theta^{(c)}$, and end iteration.
	\STATE \textbf{Reduction}\\
	$\;\;\;\;\;\;$ For all points but $\theta^{(1)}$, replace $\theta^{(i)}$ by $\theta^{(1)}+\sigma(\theta^{(i)}-\theta^{(1)})$
\ENDWHILE
\end{algorithmic}
}
\end{algorithm}

\begin{algorithm}[H]
\caption{Posterior density Maximization by Iterated Filtering}
\label{alg:mif}
{\fontsize{12}{20}\selectfont
\begin{algorithmic}
\STATE Initialise $x^{(1)}_I$ and $\theta^{(1)}$.
\STATE Unless stated otherwise, set $a = 0.975,b=2,\rho = -1/2$, and $L=round(0.75\times n)$
\FOR {$m=1$ to M}
	\STATE Sample initial conditions,  $\tilde{x}_I^{(j)}(t_0)\sim\mathcal{N}\big(x^{(m)}_I,a^{m-1}\Sigma_I\big),\;\;\;\;\;\;j=1,\dots,J$ 
	\STATE Initialise filtered states, $\tilde{x}_F^{(j)}(t_0)=\tilde{x}_I^{(j)}(t_0)$ 
	\STATE Rejuvenate parameters, $\tilde{\theta}^{(j)}(t_0)\sim\mathcal{N}\big(\theta^{(m)},ba^{m-1}\Sigma_{\theta}\big)$
	\STATE Set $\bar{\theta}(t_0)=\theta^{(m)}$
	\FOR {$i=1$ to n}
		\STATE Propagate samples, $\tilde{x}_P^{(j)}(t_i)\sim p\big(x(t_i)|\tilde{x}_F^{(j)}(t_{i-1}),\tilde{\theta}^{(j)}(t_{i-1})\big)$ 
		\STATE Compute weights, $w^{(j)}_i=p\big(y_i|\tilde{x}_P^{(j)}(t_i),\tilde{\theta}^{(j)}(t_{i-1})\big)\times p\big(\tilde{\theta}^{(j)}(t_{i-1})\big)^{\frac{1}{n}}$ 
		\STATE Draw $k_1,\dots,k_J$ such that $p(k_j=i)=w^{(i)}_i/\sum_l w_i^{(l)}$;
		\STATE and filter the predicted states $\tilde{x}_F^{(j)}(t_i)=\tilde{x}_P^{(k_j)}(t_i)$ 
		\STATE Filter the initial conditions $\tilde{x}_I^{(j)}(t_i)=\tilde{x}_I^{(k_j)}(t_{i-1})$
		\STATE Filter and rejuvenate parameters, $\tilde{\theta}^{(j)}(t_i)\sim\mathcal{N}\big(\tilde{\theta}^{(k_j)}(t_{i-1}),a^{m-1}(t_i-t_{i-1})\Sigma_{\theta}\big)$
		\STATE Set $\bar{\theta}(t_i)$ to be the sample mean of $\{\tilde{\theta}^{(k_j)}(t_{i-1})\}_{1\leq j \leq J}$
		\STATE Set $V(t_i)$ to be the sample mean of $\{\tilde{\theta}^{(j)}(t_{i})\}_{1\leq j \leq J}$
	\ENDFOR
	\STATE Set $\theta^{(m+1)}=\theta^{(m)}+V(t_1)\sum_{i=1}^n V^{-1}(t_i)(\bar{\theta}(t_i)-\bar{\theta}(t_{i-1}))$
	\STATE Set $x_I^{(m+1)}$ to be the sample mean of $\{\tilde{x}_I^{(j)}(t_{L})\}_{1\leq j \leq J}$
\ENDFOR
\end{algorithmic}
}
\end{algorithm}

\paragraph{Fast exploration of a proxy posterior density: the kMCMC algorithm}\mbox{}\\

Even when the Markov Chain is initialised close from the global model of the posterior density, the adaptation of the sampling covariance matrix $\Sigma^q$ of the pMCMC algorithm can be lengthy. We know that in the multivariate normal case, the optimal choice for $\Sigma^q$ is proportional to the covariance of the target density $p(\theta|y_{1:n})$. As the Extended Kalman Filter provides an efficient way to deterministically obtain an estimate of the likelihood $p(y_{1:n}|\theta)$ under the sde formalism (see \ref{subsub:sde}), it is natural to construct an algorithm analogous to the pMCMC, based on $\hat{p}_{EKF}(y_{1:n}|\theta)$, that will efficienctly provide an estimate of the covariance of the proxy posterior density (see Algorithm \ref{alg:kMCMC}). In addition the one or two order of magnitudes gained by estimating the likelihood with the EKF instead of an SMC algorithm, the complete absence of noise on this estimate also significantly facilitates the automatic adaptation of the \texttt{kmcmc} algorithm.

\begin{algorithm}[h]
\caption{Kalman MCMC algorithm}
\label{alg:kMCMC}
{\fontsize{12}{20}\selectfont
\begin{algorithmic}
\STATE Initialise $\theta^{(0)}$.
\STATE Use the EKF algorithm to compute  $\hat{p}_{EKF}(y_{1:n}|\theta^{(0)})$
\FOR {$i=1$ to $N^{\theta}$}
	\STATE Sample $\theta^{*}$ from $q(.|\theta^{(i)})$
	\STATE Use the EKF to compute $L(\theta^*)=\hat{p}_{EKF}(y_{1:n}|\theta^*)$ 
	\STATE Accept $\theta^{*}$ with probability $1\wedge\frac{L(\theta^{*})q(\theta^{(i)}|\theta^{*})}{L(\theta^{(i)})q(\theta^{*}|\theta^{(i)})}$
	\STATE Record $\theta^{(i+1)}$ 
\ENDFOR
\end{algorithmic}
}
\end{algorithm}

\section{\label{sec:applications}Application examples}

The following examples have been obtained by routine application of the following sequence of algorithms:

\begin{align}
\texttt{cat theta.json | ./ksimplex | ./kmcmc | ./pmcmc}\nonumber
\end{align}

They illustrate how the results presented in \cite{Dureau2013a} can be easily reproduced using \emph{SSM} to capture the time-varying drivers of epidemics. They also introduce two novel applications of epidemic modeling to increase our understanding of past epidemics and predict their future evolution and serve decision-making.

\subsection{Exploring the past: the Medieval Black Death}

The analysis of historic and recent records of plague epidemics have brought into relief 
surprising discrepancies between the characteristics of past and present epidemics that we had so far been attributing to plague. For example, the authors of \cite{Welford2009} have shown that while current laboratory-confirmed casesgenerally occur between November and April, Medieval Black Death epidemics used to burst between April and October. Have we been wrongly attributing the Black Death epidemics to the bubonic and pneumonic plagues? This question remains open, and we are simply going to illustrate here how mechanistic models could be used to provide further insight into the characteristics of present and historic epidemics.

We will be looking at two time series of 1665 epidemics in the UK, each indicating the monthly number of deaths
caused by plague in London and Eyam. Contrasting these two cases is not only interesting due to the population size difference between London  and Eyam, that had respectively 460000 and 350 inhabitants at the beginning of the epidemics, but also due to the peculiar story of the city of Eyam \cite{Race1995}. As some villagers started to die from plague, the clergyman William Monpesson decided to isolate the village in order to protect the neighbouring cities of Northern  England. During one year, Eyam sacrificed and lived in quarantine. Food was cautiously supplied so that villagers did not starve. Yet, at the end of the epidemic 250 people had died.

This story sheds a particular light on the following time series that can be found in the \emph{Bills of Mortality}.  Since 1932, these records had been filled by English doctors, who were required to monitor the deaths due to tuberculosis, small pox, measles, French pox, and plague:

\begin{figure}[h]
\begin{centering}
\includegraphics[scale=0.5]{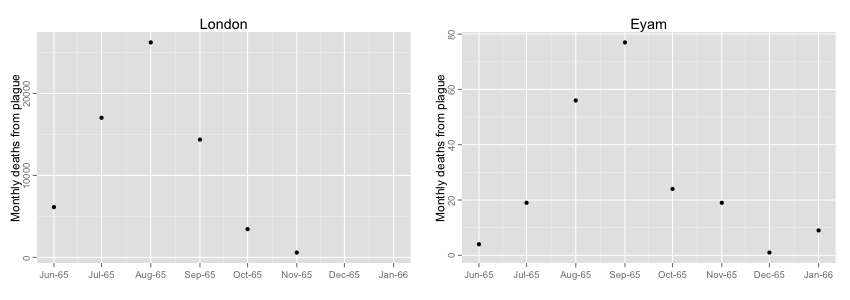}
\par\end{centering}

\caption{\label{fig:What-could-habe}Monthly number of deaths caused by Plague in London and Eyam}
\end{figure}

To analyse this dataset we use the model introduced in section \ref{sec:plague}. The timing and amplitude of seasonal forcing, as well as initial conditions, reproduction rates in each city, and life expectancy with plague, are estimated. We make no assumption
on the type of plague at stake, allowing the life expectancy after infection to lie between one and seven days
(respectively corresponding to pneumonic and bubonic plague). The resulting estimates of the transmission potential of plague in each city, as well as life expectancy with plague, are the following:

\begin{figure}[h]
\begin{centering}
\includegraphics[scale=0.5]{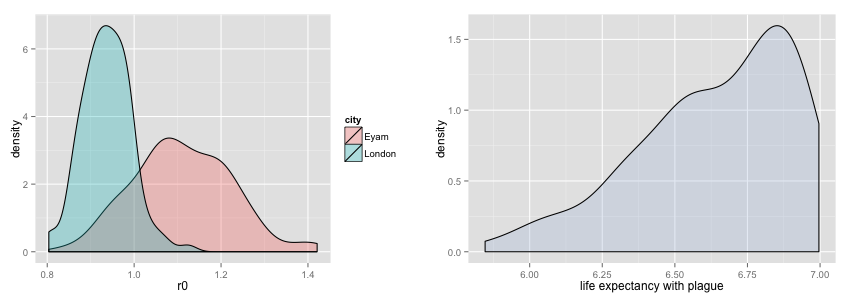}
\par\end{centering}

\caption{\label{fig:What-could-habe}Posterior densities of R0 in each city, and of the life expectancy with plague.}
\end{figure}

These results provide information that could not have been inferred from direct observation of the time series of
deaths in each city. First, they suggest that the isolation and living conditions in Eyam lead to a higher transmissibility of the
disease. Furthermore, life expectancy after infection appears to be close to one week, suggesting that this epidemic
was a bubonic plague rather than a pneumonic plague. The latter seems to be confirmed by historical records.

For the sake of transparency, and to foster further explorations of this problem and the data provided 
in \cite{Welford2009}, the following repository provides the means to easily reproduce the presented results: https://github.com/JDureau/plague-UK-1665.

\subsection{Real-time monitoring of the H1N1 transmission rate}

We have proposed in \cite{Dureau2013a}  a generic solution to monitor the transmission rate of a pathogen during an epidemic from incomplete and uncertain  measures of its spread among a population. 

The proposed methodology relies on compartmental models in which some parameters are allowed to vary over time
following a diffusion. It then helps to understand what are the underlying and unobserved causes of observed epidemic 
dynamics. We illustrated this approach on a time series of H1N1 cases recorded in London during the 2009 pandemic:

\begin{figure}[H]
\begin{centering}
\includegraphics[scale=0.7]{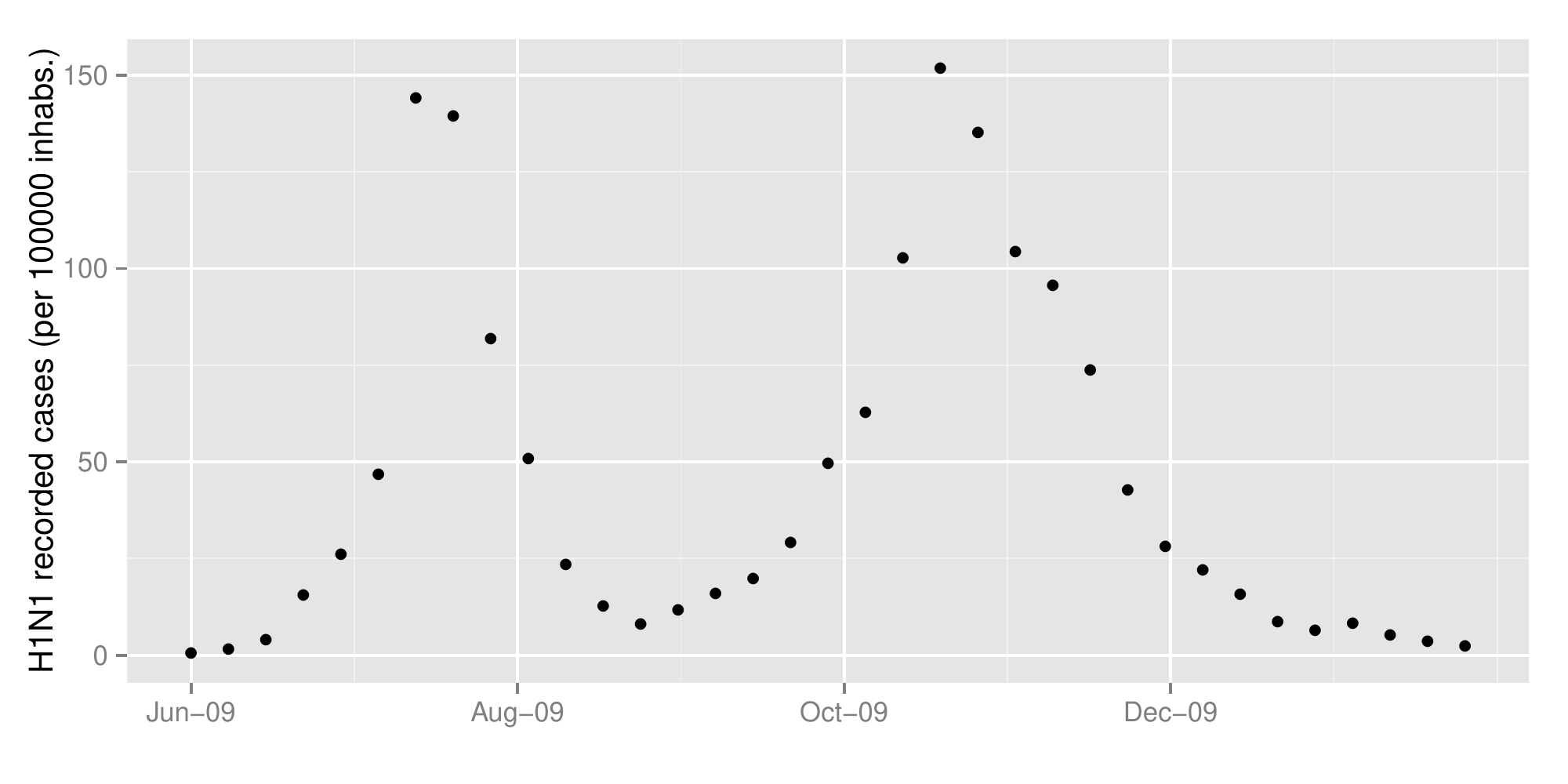}
\par\end{centering}

\caption{\label{fig:H1N1_data}Weekly number of H1N1 cases recorded in London during the 2009 epidemic (courtesy of the Health Protection Agency)}
\end{figure}

\begin{figure}[h]
\begin{centering}
\includegraphics[scale=0.5]{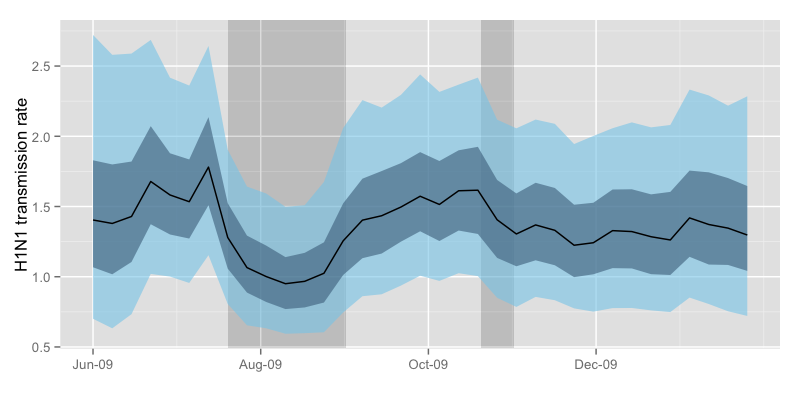}
\par\end{centering}
\caption{\label{fig:H1N1_beta}Estimated trajectory of the effective transmission rate. Darker grey areas indicate holiday periods. Light and dark blue areas respectively indicate 95\% and 50\% credible intervals}
\end{figure}

The unusual shape of the epidemic trajectory, exhibiting two peaks, reflects variations of extrinsic quantities that drive its evolution. Holidays, and their subsequent impact on the frequency at which people meet and infect  each other, provide a natural explanation to the decline of the first and second waves. However, the additional role of climate on the transmissibility of influenza is also debated, and media may have played an important role on individual awareness and behaviour. We show that according to the model we use, the effective transmission rate of H1N1 evolved in the  way illustrated by Fig \ref{fig:H1N1_beta}.

These results confirm that holidays have been the main driver of the epidemic, and further quantifies their impact on the transmission rate of influenza. For example, it shows that the impact of summer holidays is about twice more important than fall holidays, providing an indication on the potential impact of closing schools as a mean to mitigate an epidemic.

The following repository provides the means to easily reproduce the presented results: https://github.com/JDureau/H1N1-London-2009.

\subsection{How many severe cases of dengue in Madeira next year?}

Until last year, dengue had disappeared from the European continent. The last epidemic goes back to 1927-1928, 
in Greece. However, concerns of a return of dengue in Europe had started to rise in the recent years, due to the dissemination 
of \emph{Aedes albopictus} across European countries. This mosquito plays a central role in dengue transmission, as it serves as a vector for the  virus.

\begin{figure}[h]
\begin{centering}
\includegraphics[scale=0.5]{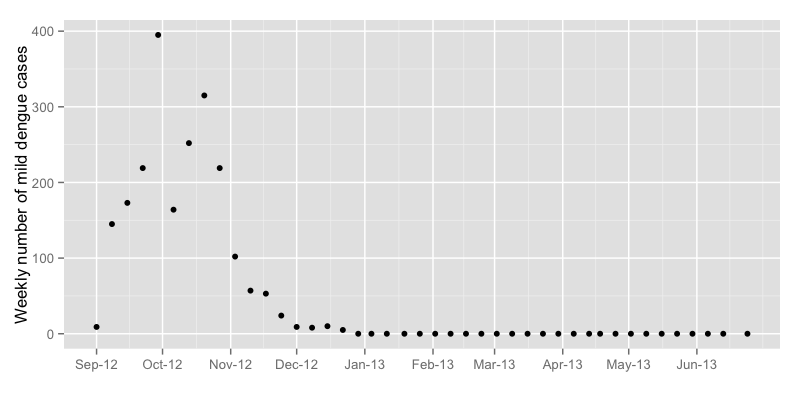}
\par\end{centering}
\caption{\label{fig:dengue_data}Weekly number of dengue cases recorded in Madeira}
\end{figure}

In September 2012, a first epidemic occured in Europe. 2159 cases were recorded over 3 months in the Portughese 
island of Madeira. Among these case, a few individuals were hospitalised for mild symptoms of fever but no severe case
has been recorded. Fig. \ref{fig:dengue_data} shows the corresponding time series, supposing that no cases have been recorded between February and July 2013, which is when this document was written.

Although multiple and crucial aspects of dengue transmission are still to be explored, some epidemiologists argue that
severe cases are more likely to correspond to secondary infections \citep{Ranjit2011}. After having previously been infected with one of
the 4 dengue strains, an individual that is re-infected with another strain would have a much higher probability of developing 
severe symptoms as hemorragic dengue fever. If we follow this assumption, and consider that all infections that occured
in 2012 were primary infection, there is a risk for severe cases in 2013 if any of the primary infected gets re-infected.

We illustrate here how mechanistic models can be used to forecast coming epidemics while reflecting the different sources of uncertainty. As described in \ref{sec:dengue}, we have extended a multi-strain model that had been introduced in \cite{Aguiar2011} to study dengue dynamics in South-East Asia.

\begin{figure}[H]
\begin{centering}
\includegraphics[scale=0.5]{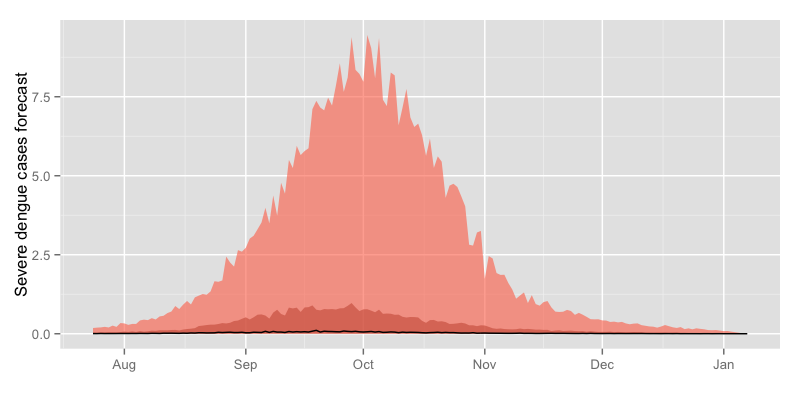}
\par\end{centering}
\caption{\label{fig:dengue_forecast}Forecasted evolution of the weekly number of sever cases in Madeira.}
\end{figure}

Under these assumptions, the data from the 2012 epidemic can be used to reconstruct the current state of immunity of 
the population of Madeira, and to project its evolution. Following a Bayesian approach allows to reflect the 
available information on the respective lengths of the infectivity and cross-immunity periods, as well as the uncertainty 
on the proportion of asymptomatics and initial state of the population immunity. We nonetheless consider that only less
than 5\% of the population had already been infected with dengue before September 2012. Accordingly, the predicted number of sever dengue cases occurring each week is illustrated in Fig. \ref{fig:dengue_forecast}.

Naturally, these preliminary results shall be explored further to strengthen the evidence provided by this analysis. To that end, the following repository provides the means to easily reproduce the presented results: https://github.com/JDureau/dengue-Madeira-2012.

\bibliographystyle{apalike} 

\bibliography{Biblio}

 \end{document}